\documentclass[onecollarge,natbib]{svjour2}
\bibpunct{[}{]}{;}{n}{}{,} 
\smartqed  
\usepackage{graphicx,amsmath}
\journalname{Few-Body Systems (EFB22)}
\def\be{\begin{eqnarray} &&}

\def\ee{\end{eqnarray}}
 \usepackage{wrapfig}
 \usepackage[normalem]{ulem}
 \usepackage{graphicx,color}
 \usepackage{enumerate}

     \font\tenbifull=cmmib10 scaled 1200 
     \font\tenbimed=cmmib9
     \font\tenbismall=cmmib7
\mathchardef\bbkappa="7114
\mathchardef\bbrho="711A
\mathchardef\bbsigma="711B
\mathchardef\bbtau="711C
\mathchardef\bbvarrho="7125
\mathchardef\bbvarsigma="7126
\mathchardef\bbxi="7118

\def\be{\begin{eqnarray} &&}

\def\ee{\end{eqnarray}}

\makeindex

\def\beq{\begin{equation}}
\def\eeq{\end{equation}}
\usepackage{amsbsy,amsmath}

\newcommand{\bq}{\begin{eqnarray}}
\newcommand{\eq}{\end{eqnarray}}

\begin{document}

\title{ Solutions of the Bethe-Salpeter 
Equation in
Minkowski space: a comparative study} 




\author{Giovanni Salm\`e  \and Tobias Frederico \and Michele Viviani
 }


\institute{
          Giovanni Salm\`e \at
             INFN, Sezione di Roma, Italy
	       \and
          Tobias Frederico \at
           Dep. de F\'\i sica, ITA, S\~ao Jos\'e dos
Campos, S\~ao Paulo, Brazil
                   \and
 Michele Viviani \at
 INFN, Sezione di Pisa, Italy 
}
\date{}
\maketitle
\begin{abstract}
The Bethe-Salpeter   equation for a
two-scalar, S-wave 
bound system, interacting through a massive scalar, is investigated within the
ladder approximation. By assuming  a Nakanishi integral 
representation of the Bethe-Salpeter amplitude, one can deduce  new integral 
equations that can be solved and quantitatively studied, overcoming the analytic
difficulties of the Minkowski space.
 Finally, it is shown that the
Light-front distributions of the valence state,   directly obtained
from the Bethe-Salpeter amplitude,  open  an effective window  for 
studying the
two-body  dynamics.

\end{abstract}

\section{Introduction}
\label{intro}
Recently, there has been a renewal of interest in looking for solutions of the 
 Bethe-Salpeter Equation (BSE) in Minkowski space (see, e.g., Refs. \cite{CK1,CK2,FSV1,FSV2}
and, for  earlier studies,  see Ref. \cite{KW}),
since the perspective of establishing  a new  framework where  both bound and scattering 
states  can be investigated in a non
perturbative regime is very attractive (let us remind that an
integral equation is a non perturbative tool). As is well known, to get
solutions in Minkowski space represents a big challenge,
since the analytic structure of the relevant quantities, like the irreducible kernel present
 in BSE and particularly the Bethe-Salpeter
amplitude (BSA), has to be explicitly considered.
A possible way to overcome such a difficulty is provided  by  assuming that  the
BSA have an expression inspired by   the perturbation-theory integral 
representation (PTIR), introduced by N. Nakanishi in the sixties (see  Ref.\cite{nak71} for
the textbook presentation of the topic) for describing   N-leg transition
amplitudes. As emphasized  by the name itself of the approach,
the Nakanishi integral representation holds in a perturbative regime.
For the BSA,  one should  consider the expression 
corresponding to the 3-leg transition amplitude.
Hence the question if and to what extent actual solutions of the BSE, that have a non perturbative 
nature, can be
expressed through the Nakanishi PTIR.

In the present contribution,  we give a brief presentation (see Refs.
\cite{FSV1,FSV2} for details) of the application of  the Nakanishi approach 
for solving the  ladder BSE, within a non-explicitly covariant Light-front (LF)
framework. The system analyzed is composed by   two massive  
scalars, interacting through a massive scalar, in a bound S-wave state.
Let us shortly recall the BS approach
(see, e.g., \cite{zuber} for details).

The 4-point Green's function (FPGF), 
$
G(x_{1},x_{2};y_{1},y_{2})=<0\,|\,
T\{\phi_{1}(x_{1})\phi_{2}(x_{2})\phi_{1}^{+}(y_{1})
\phi_{2}^{+}(y_{2})\}\,|\,0>$, 
 fulfills an integral equation:  
$G=G_0~+~G_0~{\cal I}~G$, 
where  $G_0$ is the  Green's function of two free particles and 
 ${ \cal I}$  the  irreducible kernel, 
  given by the 
infinite sum of  Feynman graphs,  that cannot be
disconnected after cutting along one line the scalar propagations, but  without
touching the exchanged-scalar propagations. 
All the expected contributions can be obtained by iterating the previous
integral equation.
By inserting a complete Fock basis  in the expression of the FPGF
 and moving to the Fourier space,  one  
can single out the bound state contribution
(assuming only one non-degenerate bound state, for the sake of simplicity). Such
a contribution to the FPGF
appears as a pole, viz.
\be G_b(k,q;p_b)\simeq {i\over (2\pi)^{-4}}~
\frac{\phi(k;p_b)~\bar{\phi}(k;p_b)}
{2\omega_b(p_{0}-\omega_b+i\epsilon)}
\ee
where   $\omega_b=\sqrt{M^2_b+|{\bf p}_b|^2}$, $p^2_b=M^2_b$ and  $\phi(k;p_b)$ 
 is  the Fourier
 transform of the  BSA for a bound state,
given by
$
\langle0|T\{\phi_{1}(x_{1})\phi_{2}(x_{2})\}
|p_b\,\beta\rangle$.
For $p_0\to \omega_b$ the FPGF is approximated by
$G\simeq G_b+~regular~terms$
and one deduces from $G=G_0~+~G_0~{\cal I}~G$
the integral equation determining the BSA for a bound
state, viz
\be   \phi(k;p_b,\beta)= G_0(k;p_b,\beta) ~\int
d^{4}q'~{\cal I}(k,q';p_b)~\phi(q';p_b,\beta) \ee
where
$G_0= i^2/\{[(p_b/2+k )^2-m^2+i\epsilon]~[(p_b/2-k)^2-m^2+i\epsilon]\}
$ with $p_b=p_1+p_2$ and $k=(p_1-p_2)/2$.
Notably, the same irreducible kernel, ${\cal I}(k,q';p_b)$, present in the
integral equation for the FPGF,   is acting.

\section{The Nakanishi integral representation of the BS amplitude}
\label{sec:1}
The  PTIR $N$-leg transition amplitude can be formally written as
\be f_N(s)\sum_{\cal G} ~ f_{\cal G}(s)\propto~\prod_h\int_0^1dz_h
\int_0^\infty 
d\gamma 
\frac{\delta(1-\sum_h z_h)~\phi_N(z,\gamma)}{(\gamma-\sum_h z_hs_h)} \ee
where ${\cal G}$ indicates a generic Feynman diagram contributing to the
$N$-leg amplitude, and   
$ {\phi}_N(z,\gamma)=\sum_{\cal G}  \tilde{\phi}_{\cal G}(z,\gamma) $. For the
$3$-leg case, the so-called vertex function, one has 
\be f_3(s)=\int_{-1}^1 dz 
\int_{-\infty}^\infty d\gamma \frac{\phi_3(z,\gamma)}{\gamma-{p^2\over 4}- k^2-z k\cdot p-i\epsilon}
\ee
Within the BS framework, can such an elegant expression  be exploited
for modeling the BSA (see, e.g.,\cite{KW,CK1,CK2,FSV1,FSV2})?   
In particular, if the BSA for both bound and scattering states \cite{FSV1} has
actually such a form, then one can determine the Nakanishi weight function,
$\phi_3$, through an integral equation (indeed one can construct two integral
equations as explained in what follows), that  can be directly deduced from 
the BSE. This achievement allows one to answer  the question: 
 can the Nakanishi representation of the vertex function, elaborated in
perturbation theory,  be used in a non perturbative realm,
  as the BS framework does (recall that BSE is  
  an integral equation)? 
  
  A suitable integral equation for the relevant Nakanishi weight function can be obtained by applying the
  so-called LF projection method (see, e.g., \cite{FSV1,FS} and references
  quoted there
  in). This  amounts to
integrate the BSE on the LF variable $k^-=k^0+k_z$,   but in order to perform 
the needed 
integration one has to
know the analytic structure of the unknown BSA.
 Let us take the Nakanishi vertex function as an 
 Ansatz for the BS 
 amplitude, $\Phi_b(k,p)$ 
 and  then integrate it on the
  $k^-$. As a result,  then 
 one gets the 
 valence component of the
 state of the interacting system (after expanding it onto the Fock basis)
 \cite{CK1,FSV1}
\be \psi_{n=2}(\xi,k_\perp)={p^+\over \sqrt{2}}~\xi(1-\xi)
\int {dk^- \over 2 \pi}
\Phi_b(k,p)
={\xi(1-\xi)\over \sqrt{2}}\int_0^{\infty}d\gamma'~\frac{
g_b(\gamma',1-2\xi;\kappa^2)}
{[\gamma'+k_\perp^2 +\kappa^2+\left(2\xi-1\right)^2 {M^2_b\over 4}-i\epsilon]^2}
\label{val}\ee
where $g_b$ is the Nakanishi weight function (adopting the standard notation of
Refs. \cite{FSV1,FSV2}, see also \cite{CK1,CK2} for the explicitly covariant
LF approach).
  Applying the LF projection to both sides of the BSE, one obtains
\be \int_0^{\infty}d\gamma'\frac{g_b(\gamma',z;\kappa^2)}{[\gamma'+\gamma
+z^2 m^2+(1-z^2)\kappa^2-i\epsilon]^2} =
\int_0^{\infty}d\gamma'\int_{-1}^{1}dz'~V^{LF}_b(\gamma,z;\gamma',z')
g_b(\gamma',z';\kappa^2).
\label{eq1} \ee
where $\kappa^2=m^2-M^2_b/4$, and 
  $ V^{LF}_b(\gamma,z;\gamma',z')$ is determined  by the irreducible kernel 
  ${\cal I}(k,k',p)$. Moreover, if one assumes that 
the uniqueness theorem of the Nakanishi weight function
(see  Ref. \cite{nak71} for the demonstration within the PTIR framework)
holds also in the non perturbative regime,  one can deduce from Eq. (\ref{eq1})
 a simpler integral equation for  the weight function, viz 
\be g_b(\gamma,z;\kappa^2)=~\int_{0}^{\infty}d\gamma'\int_{-1}^{1}dz'\;
{\cal V}_b(\gamma,z;\gamma',z';\kappa^2)
g_b(\gamma',z';\kappa^2)\label{eq2}\ee
where ${\cal V}_b(\gamma,z;\gamma',z';\kappa^2)$ is a new kernel, properly 
related to 
$V^{LF}_b(\gamma,z;\gamma',z')$ (see \cite{FSV1,FSV2}) .

\begin{figure}
\parbox{9cm}{\includegraphics[width=9.cm]{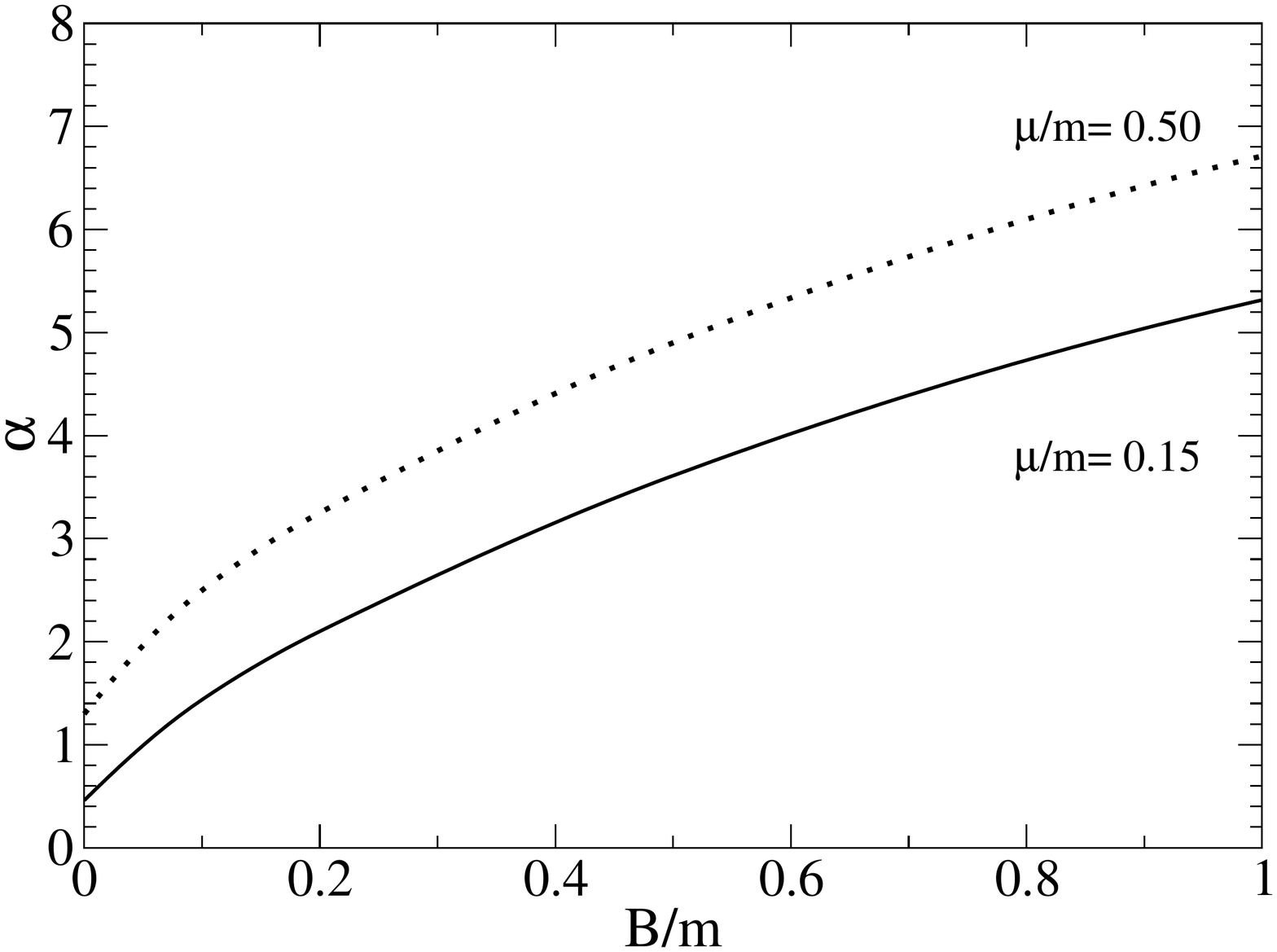}} \hspace{-1.0cm}
\parbox{9cm}{\includegraphics[width=9.0cm]{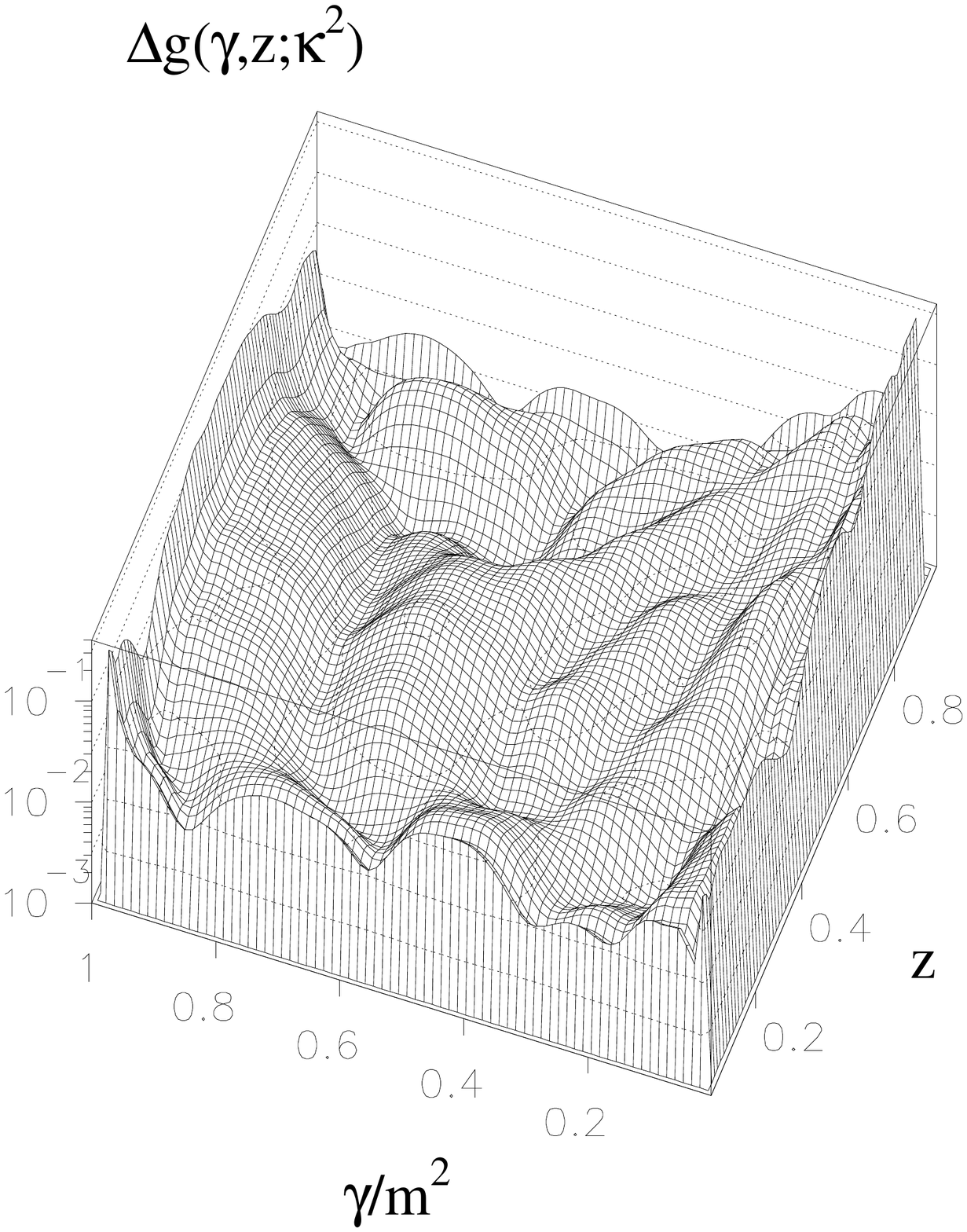}}
\vspace*{-1.3cm}
\caption{Left panel: Values of the coupling constant $\alpha=g^2/(m^2 16 \pi)$ vs the
binding energy $B/m$ for two values of the exchanged scalar mass
$\mu/m=0.15,~0.50$.  Right panel: $\Delta g(\gamma,z;\kappa^2)=~\left|g_U(\gamma,z;\kappa^2)-
g_V(\gamma,z;\kappa^2)\right|$ for$B/m=1.0$ and $\mu/m=0.5$.}
\end{figure}
\section{Results and conclusions}
The two integral equations, (\ref{eq1}) and (\ref{eq2}), have been carefully
investigated, in ladder approximation, in Ref. \cite{FSV2}, and the numerical results have been 
compared to the
ones in Ref. \cite{KW} (where the  uniqueness 
theorem
was exploited, but with standard variables in Minkowski space) and in Ref.
\cite{CK1}. The quantitative studies proceed by assigning a value to the 
binding
energy of the system, $B/m=2-M_b/m$, 
and then determining i) the coupling constant $\alpha=g^2/(m^2 16\pi )$ (recall
that the interacting Lagrangian is
${\cal L}=g\phi^2
\chi$) and the corresponding Nakanishi weight function. The values of 
$\alpha$, for different values of the exchanged scalar mass $\mu/m$, agree
 very
accurately with the ones in Refs \cite{KW,CK1}, and they are illustrated in Fig.
1, left panel (for a detailed analysis see Ref. \cite{FSV2}). In the right panel
of Fig. 1, it is presented for $B/m=1.0$ and $\mu/m=0.5$ the difference between
the Nakanishi weight function obtained from Eq.  (\ref{eq1}),
shortly indicated by $g_V$, and the one obtained from Eq.  (\ref{eq2}), $g_U$. 
In general, the differences are quite small (for more comparisons see Ref.
\cite{FSV2}), and it should be reminded that $g_b(\gamma,z;\kappa^2)$ must vanish
at the end-points $z=\pm 1$ 
and  quickly 
fall off for increasing $\gamma/m$.
From the valence wave
function, Eq. (\ref{val}), one can calculate \cite{FSV2} i) the probability of the valence
component, $P_{val}$, ii) the LF distribution of the 
longitudinal-momentum fraction, $\phi(\xi)$,
 and  iii) LF distribution of the 
transverse momentum, ${\cal P}(\gamma)$, given by 
\be
\phi(\xi)={1 \over (2 \pi)^3}  {1\over 2~\xi(1-\xi)}~ \int 
 d{\bf k}_\perp~
 \psi^2_{n=2/p}(\xi,k_\perp) ~, ~  {\cal P}(\gamma)= {1 \over 2(2 \pi)^3}  \int_0^1 {d\xi\over 2~\xi(1-\xi)}~  
 \int_0^{2\pi} d\phi
 \psi^2_{n=2/p}(\xi,k_\perp)\ee
Those distributions, shown in Fig. 2 for some values of 
$B/m$
and $\mu/m=0.5$, open an interesting window on the
internal dynamics of the system. In particular, from the present, preliminary
analysis, the $\gamma/m$ tail of the
transverse-momentum distribution is governed by the ladder kernel, providing  a
power-like behavior. This will be thoroughly investigated elsewhere \cite{FSV3}.

In conclusion, combining the Nakanishi integral representation of the BS
amplitude and the LF projection one can afford the investigation of the BSE in
Minkowski space. In perspective, this can allow to develop applications in
different areas, from solid state physics to hadronic physics.
Calculations are in progress for i) the scattering length and ii) 
the cross-box contribution with the uniqueness theorem.

\begin{figure}
\parbox{9cm}{\includegraphics[width=9.cm]{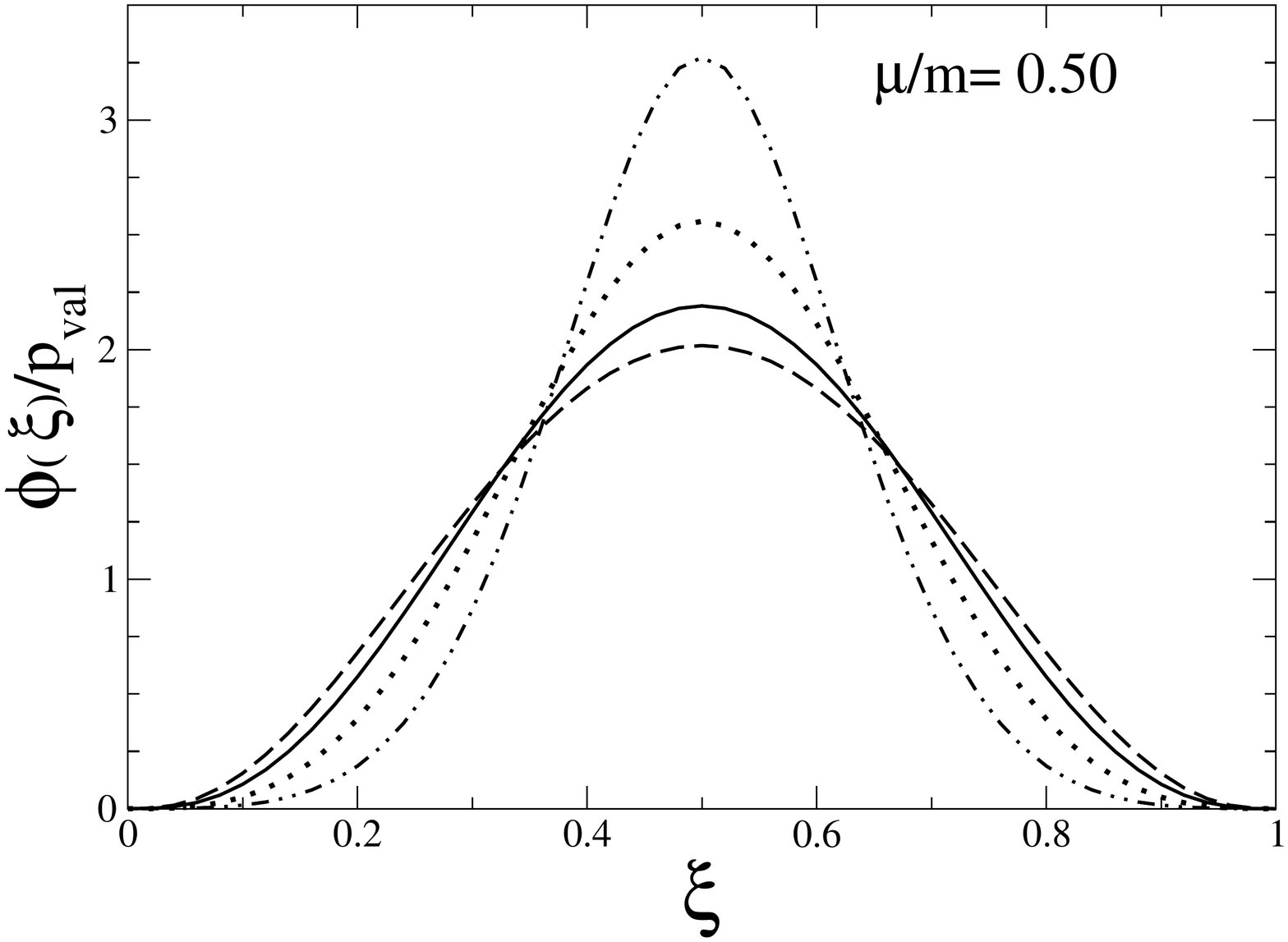}}\hspace{-0.5cm}
\parbox{9cm}{\includegraphics[width=9.cm]{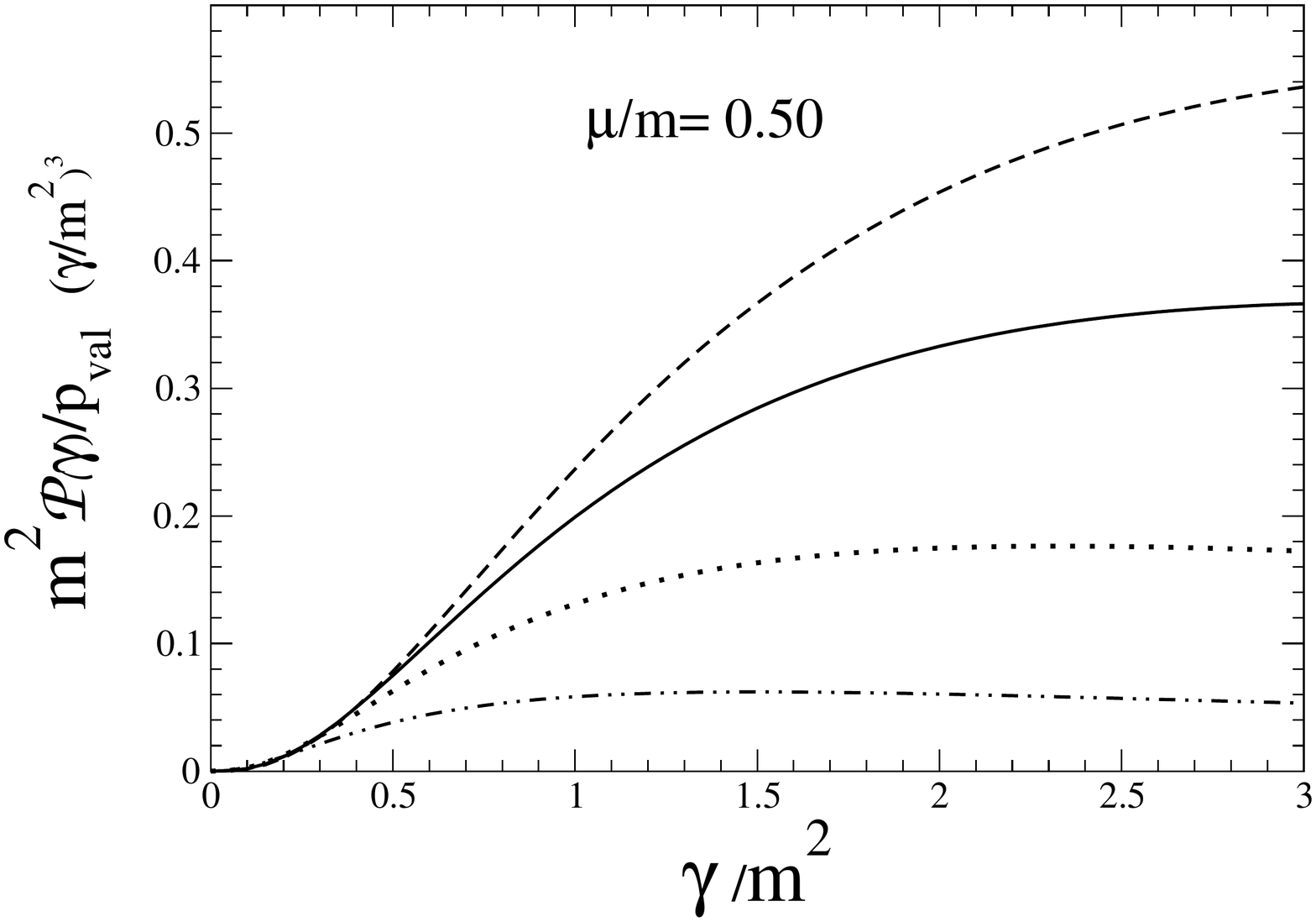}}
\caption{Left panel: the LF longitudinal-momentum fraction distribution, $\phi(\xi)$,  
 vs the longitudinal-momentum fraction 
$\xi=k^+/M$.  Dash-double-dotted line: $B/m=0.20$. Dotted line: $B/m=0.50$. 
Solid line: $B/m=1.0$. Dashed line: $B/m=2.0$.
(N.B. $\int^1_0 d\xi~\phi(\xi)= P_{val}$).(After from   Ref.\cite{FSV2}) Right panel:
the same of the left panel but for the  LF transverse-momentum distribution, 
${\cal P}(\gamma)~(\gamma/m^2)^3$ vs $\gamma/m^2$ (recall that
$\gamma=k_\perp^2$). 
(Adapted from   Ref.\cite{FSV2}).}
\end{figure}

\medskip
T.F acknowledges the partial financial 
support from 
 the Conselho Nacional
de Desenvolvimento Cient\'{\i}fico e Tecnol\'ogico (CNPq),
the  Funda\c c\~ao de Amparo \`a Pesquisa do
Estado de S\~ao Paulo (FAPESP)

\end{document}